# Exploring Wetting and Optical Properties of CuAg Alloys via Surface Texture Morphology Analysis

Krzysztof Wieczerzak[a], Grzegorz Cios[b], Piotr Bała[b,c], Johann Michler[d], Benedykt R. Jany[e*]

[a]Department of Materials Science, Faculty of Mechanical Engineering and Aeronautics, Rzeszow University of Technology, al. Powstańców Warszawy 12, 35-959 Rzeszow, Poland
[b]Academic Centre for Materials and Nanotechnology, AGH University of Krakow, al. A Mickiewicza 30, 30-059 Krakow, Poland
[c]Faculty of Metals Engineering and Industrial Computer Science, AGH University of Krakow, al. A Mickiewicza 30, 30-059 Krakow, Poland
[d]Laboratory for Mechanics of Materials and Nanostructures, Empa, Swiss Federal Laboratories for Materials Science and Technology, CH-3602 Thun, Switzerland
[e]Marian Smoluchowski Institute of Physics, Faculty of Physics, Astronomy and Applied Computer Science, Jagiellonian University, PL-30348 Krakow, Poland

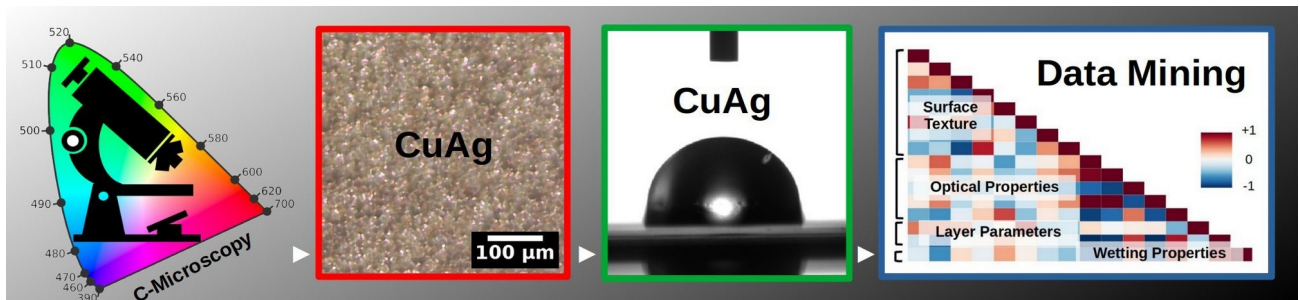

**Abstract**

Copper–silver (CuAg) alloys are increasingly explored for applications in high-performance electrical and electronic systems, owing to their unique combination of high electrical and thermal conductivity and enhanced mechanical strength. Nevertheless, a thorough understanding of how these alloys surface characteristics fundamentally influence properties remains largely underdeveloped. Here, we explored the complex interplay between surface texture morphology, layer composition, wetting, and optical properties of Cu, Ag, and CuAg thin films deposited on textured silicon substrates via magnetron sputtering. Employing data mining and machine learning techniques, we identified robust correlations between contact angle and surface fractal dimension across all layer types promoting Cassie-Baxter surface state formation. Our analysis revealed a significant connection between layer thickness and surface topography entropy deficit, suggesting a dynamic evolution of surface order/disorder during metal film growth. Furthermore, we observed that contact angle sensitivity to layer thickness implied a correlation with microstructure evolution. Through K-Means clustering, we successfully categorized the formed surface textures morphology. Finally, a Random Forest regression model was developed to accurately predict water contact angles (Mean Absolute Error around 5 deg) using only texture and optical parameters. The model, along with accompanying Python code, is publicly available. Our findings establish a pathway towards targeted surface texture morphology engineering for tailored material performance.

---

\* Corresponding author e-mail: benedykt.jany@uj.edu.pl

# Introduction

Copper-silver (CuAg) binary alloys are attracting increasing attention due to their advantageous combination of high electrical conductivity and mechanical strength. This makes them well-suited for a range of demanding high-performance applications[1,2,3,4]. CuAg alloys are particularly noted for their superior electrical conductivity, a key requirement in applications such as electrical connectors and components operating in high-conductivity environments[5,6]. Furthermore, CuAg alloys exhibit good thermal conductivity, facilitating effective heat dissipation in electronic devices[7]. Although not among the strongest metallic materials overall, CuAg alloys offer a favorable trade-off between electrical conductivity and mechanical strength. Their hardness and tensile strength can be substantially increased relative to pure copper or silver, depending on the chemical composition, phase structure, and thermo-mechanical treatment. Cold deformation and directional solidification techniques can further improve their mechanical performance[8]. Notably, the addition of silver results in the formation of nano-sized Ag precipitates, contributing to strengthening through solid-solution and precipitation hardening mechanisms[9]. Incorporating zirconium (Zr) into CuAg alloys produces multicomponent materials, including CuAgZr metallic glasses (MGs). The amorphous structure of these MGs results in distinct combinations of strength, ductility, and viscoelasticity – characteristics absent in conventional crystalline alloys[10,11]. Nevertheless, a thorough understanding of how these alloys' surface characteristics fundamentally influence their properties remains largely underdeveloped. Examining the relationship between surface texture morphology, like roughness and contact angle, a measure of wettability, is essential[12,13,14,15,16,17]. Different textures (square, circular, hemispheric, and triangle) processed on metal alloy surfaces exhibit varying structural accuracy and surface morphology. Square textures demonstrate the highest accuracy and contact angle, while triangle textures display the worst formation quality due to the stepping effect mechanism[18]. The surface texture morphology also influences the optical properties of metal alloys. Surface texture morphology affects optical characteristics including roughness, glossiness, color, and reflectivity[19]. Furthermore, surface texturing can enhance tribological performance, such as increased fatigue strength, corrosion resistance, wear resistance, anti-biofouling hydrophobicity, and load-carrying capacity[20]. It was observed that for the CuAg alloys the surface morphology and film structure were highly dependent on both the alloy composition and the annealing conditions[21]. Despite significant advancements in metal alloys, a comprehensive understanding of how surface textures – including roughness and wettability – fundamentally impact their properties, optical characteristics, and tribological performance remains a critical and largely unexplored area of research. We systematically investigated the influence of surface texture morphology, layer composition, wetting, and optical properties of copper (Cu), silver (Ag), and copper-silver (CuAg) films deposited on textured silicon substrates via DC sputtering. Utilizing data mining and machine learning techniques, promising tools in microscopy and material science[22],[23],[24], we identified robust correlations between contact angle and surface texture morphology properties, such as surface fractal dimension. We also uncovered universal relationships governing the behavior of all CuAg phases and observed a significant link between layer thickness and surface entropy deficit. Employing K-Means clustering, we categorized the resulting surface textures. Furthermore, a Random Forest regression model was developed to predict surface wetting properties – specifically, water contact angles – based solely on texture and optical parameters. The model and associated Python code and data are publicly available via Zenodo repository[25]. Our studies offer a pathway towards targeted surface texture morphology engineering.

**Results**

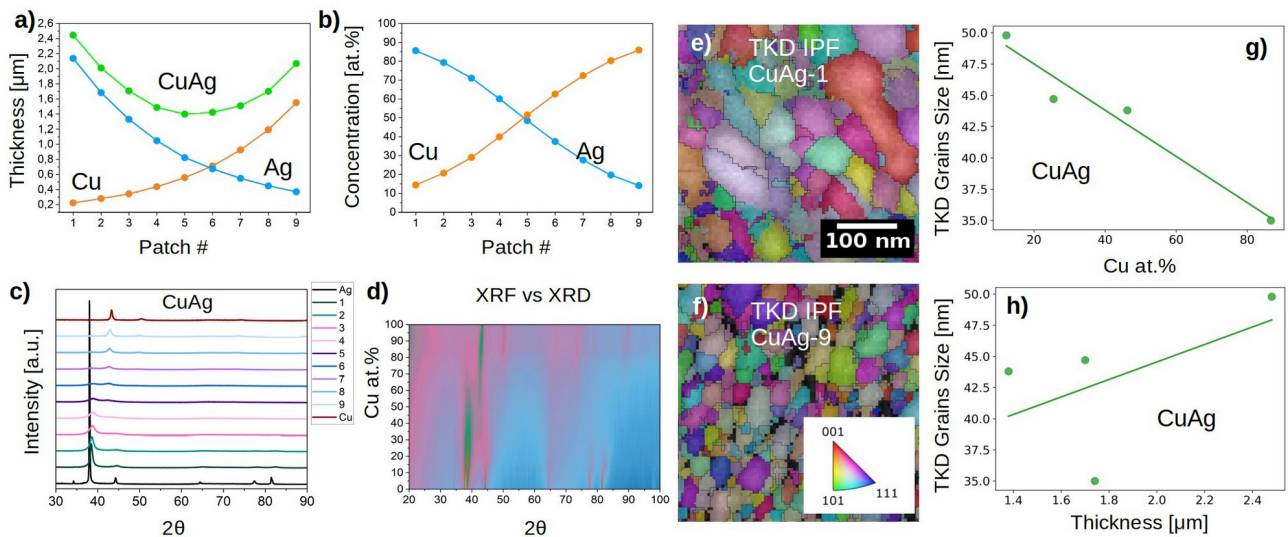

*Figure 1: Synthesized CuAg samples fabricated as layers on a textured Si substrate i.e. Cu, Ag and CuAg layers. a) Layer thickness and b) composition are shown as a function of sample number (Patch), as determined by XRF measurements. c) XRD diffractograms (intensity vs. 2θ) for various CuAg samples. d) A two-dimensional plot of the diffractograms for different copper concentrations. It is see that the synthesized samples have a copper concentrations from 0-100 at.%. e)-f) Exemplary SEM transmission Kikuchi diffraction (TKD) Inverse Pole Figures (IPF) of CuAg-1 and CuAg-9 samples respectively. g) Crystallographic grains size (from TKD) as a function of copper concentration in CuAg. It is see that the grains size decreases with increasing copper content. h) Crystallographic grains size (from TKD) as a function of CuAg layers thickness. It is see that the grains size increases also with increasing layer thickness.*

Figure 1 presents the results of the characterization of 27 synthesized PVD (DC Sputtering) CuAg samples with layers deposited on surface textured Si substrate – specifically, Cu, Ag, and CuAg layers. The samples were characterized for thickness, as determined by XRF, which varied from approximately 0.2 to 2.4 microns Fig. 1a). The XRF-determined Cu at.% concentration for the CuAg alloy layers ranged from 10 to 90 at. % Fig. 1b) (see also Fig. S1 and Table S1 in Supporting Informartion). Fig. 1c)–d) shows the evolution of the diffraction patterns as the copper content increases, reflecting the structural transition from pure Ag to pure Cu. The (111) diffraction peak of silver, initially centered at ~38.15°, progressively broadens and shifts toward higher 2θ angles with increasing Cu concentration. This broadening likely arises from a combination of factors, including solid solution formation, microstrain, and crystallite size reduction. Around 60 at.% Cu, the Ag (111) peak diminishes and eventually disappears, coinciding with the emergence of the Cu (111) reflection at ~43.35°, consistent with the expected phase evolution based on the Cu-Ag binary phase diagram[26]. The samples cross sections were also evaluated using SEM transmission Kikuchi diffraction (TKD) to determine the crystallographic grain properties (see also Fig. S9 in Supporting Information). Exemplary TKD inverse pole figures (IPF) of the CuAg-1 and CuAg-9 samples are shown in Fig. 1e) and f), respectively. The grain size was determined for the CuAg alloys as the equivalent circle diameter, and the area-weighted mean was calculated. We observe that grain sizes decrease with increasing copper content, see Fig. 2g). Additionally, as the layer thickness increases, the grain sizes also increase, see Fig. 2h), as typically for the sputter deposited thin films[27]. Next, the samples were characterized by colorimetric microscopy (C-Microscopy) to determine their surface texture morphology, colorimetric, and optical properties, see Fig. 2. Exemplary C-Microscopy images of Ag, Cu, CuAg, and Si surfaces are presented in Fig. 2a), see also Supporting Information Fig. S1. The quantitative color of the samples is presented on the CIE1931 xy chromaticity diagram (D65 illuminant, 2° standard observer) in Fig. 2b). It was observed that the samples formed groups according to their phase (composition). Increasing the sample thickness resulted in an increase in optical reflectance (R) for each sample phase separately as in Fig. 2c). Since metal crystallographic grains significantly affect optical reflectance through scattering and absorption[28], this effect is primarily due to the increasing grain size with increasing layer thickness, see Fig. 1h). Simplifying, each grain functions as a micro-mirror, reflecting light. As the layer thickness increases, the reflection efficiency also improves, owing to the larger grain size and a corresponding reduction in the number of grain boundaries. Consequently, optical reflectance increases. A similar behavior was previously observed in silver thin films[29],[30]. Grain size significantly influences the imaginary part of the dielectric function and the extinction coefficient, with smaller grains showing deviations from bulk values, particularly impacting optical properties.

The refractive index decreases with increasing grain size, leading to higher reflectivity. In consequence, larger grains generally result in higher visible reflectance. A linear increase was also observed between the colorimetric dominant wavelength and copper concentration for the CuAg alloy samples Fig. 2d).

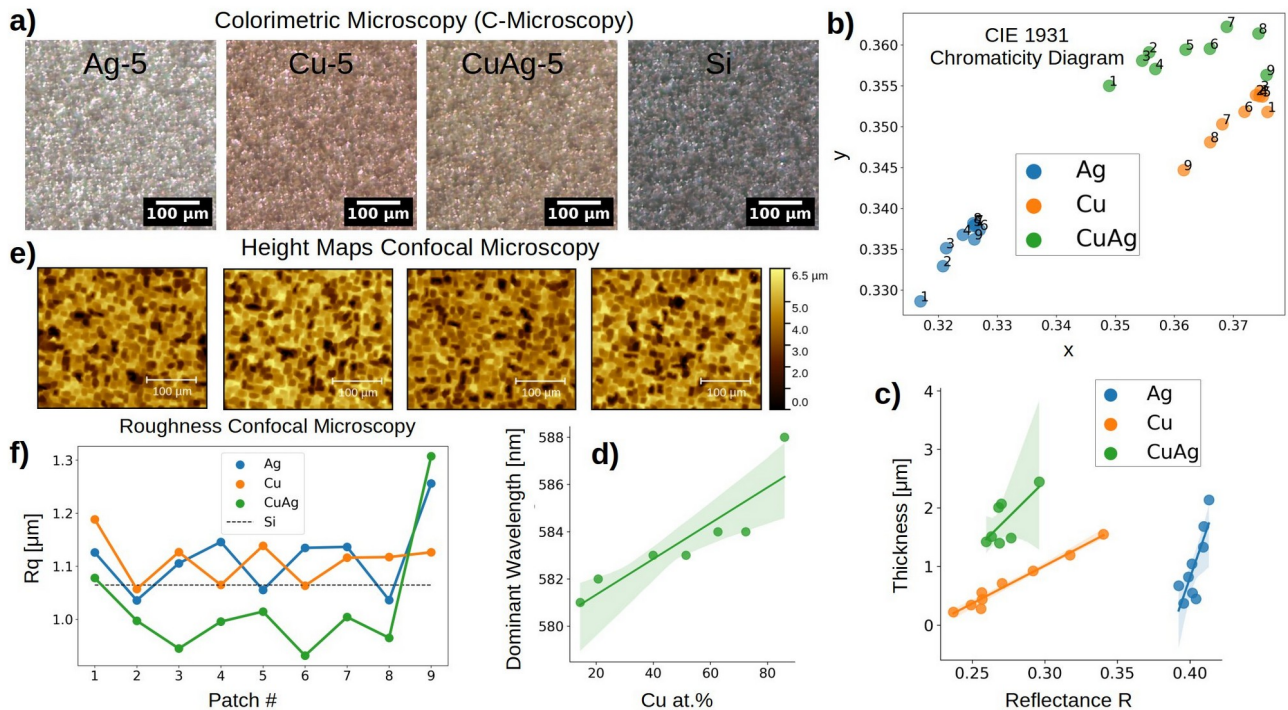

Figure 2: Results of CuAg samples imaging and analysis using Colorimetric Microscopy (C-Microscopy). a) Exemplary C-Microscopy images of Ag, Cu, CuAg, and Si surfaces. b) CIE1931 chromaticity diagram (D65 illuminant, 2° standard observer) displaying the sample positions. c) Thickness versus optical reflectance (obtained from C-Microscopy) for synthesized samples – a linear dependence is observed for each sample phases. d) Colorimetrically determined dominant wavelength versus copper concentration for CuAg layers, demonstrating a linear relationship. e) Exemplary height maps of Ag, Cu, CuAg, and Si surfaces as measured by confocal microscopy. f) Determined root-mean-square (rms) roughness (Rq) for different CuAg samples as a function of sample number (Patch). The roughness of CuAg alloys is consistently lower than that of the Si substrate, suggesting that the CuAg alloy spreads more homogeneously than Ag and Cu.

In addition, height maps were examined using confocal microscopy, with representative maps of Ag, Cu, CuAg, and Si surfaces shown in Fig. 2e). From the confocal microscopy measurements, we determined the surface roughness, i.e., RMS roughness (Rq). The Rq roughness values for different CuAg samples as a function of sample number (Patch) are presented in Fig. 2f). The roughness of CuAg alloys is consistently lower than that of the Si substrate, suggesting that the CuAg alloy spreads more homogeneously and smoothly on the Si substrate than pure Ag and pure Cu.

We used collected colorimetric microscopy (C-Microscopy) surface images to extract information about the surface texture morphology. Using free software Gwyddion, we extracted eight standard parameters describing surface texture morphology: Entropy – a measure of the uncertainty or randomness; Entropy deficit – the difference between the maximum possible entropy (for a Gaussian distribution with the same RMS value) and the estimated entropy; Fractal Dimension – a measure of surface complexity, Autocorrelation length (Sal) – a measure of the scale of the surface's texture; Autocorrelation Fastest decay direction – the direction of the fastest decay in the autocorrelation function; Autocorrelation Slowest decay length – the slowest length at which the autocorrelation function drops; and Autocorrelation Texture aspect ratio (Str) – the ratio of the fastest to slowest decay lengths of the autocorrelation, these were extracted from the 2D autocorrelation function. We also evaluated the wetting properties of each sample by measuring the water contact angle (CA), see exemplary measurements in Fig. 3a) and Fig. 3f). The average contact angles for our textured surfaces are CA(Ag) = 105°, CA(Cu) = 90°, CA(CuAg) = 94° and for comparison for our textured silicon substrate CA(Si) = 51°. Notably, the contact angles for the polished pure metal surfaces are lower[31]: CA(Ag) = 64°, CA(Cu) = 72° and for flat silicone[32] for comparison CA(Si) = 21°, indicating that surface texturing and deposition conditions significantly enhances surface water hydrophobicity.

Later, we used a data mining technique – a correlation matrix analysis – to find relationships between surface texture morphology, optical properties, layer parameters, and wetting properties. We utilized the standard Pearson correlation coefficient together with a p-value analysis to identify statistically significant correlations ($p<0.05$), see Supporting Information Fig. S2-S5. This allowed us to identify the following key statistically significant correlations for the wetting properties. The analysis identified correlations between contact angle and: Ag layer excitation purity Fig. 3b), the slowest decay length observed in Cu layers Fig. 3c), and copper concentration in CuAg layers Fig. 3d). The decrease in the water contact angle with increasing copper content can be attributed to changes in the surface texture properties with copper content; for a flat, polished CuAg alloy surface, one would expect the opposite trend i.e., an increase in contact angle with increasing copper content (see Table S2 and Fig. S10 in Supporting Information). The performed analysis also identified a correlation between the contact angle and the surface fractal dimension across all layer types Fig. 3e). This suggests that the surface becomes more hydrophobic (water-repellent) as its complexity (fractal dimension) increases. As the Fractal Dimension increases the surface is more irregular, rough, and complex, it is getting more self-similar at different scales with more surface asperities and micro/nanostructures. A higher fractal dimension implies that the surface possesses a larger effective area and more intricate micro and nanoscale features, which can more effectively trap air pockets between the asperities. When a droplet lands on such a surface, it can sit on top of these structures, with air trapped underneath. This prevents the droplet from fully contacting the solid and reducing the effective solid-liquid contact area, the contact angle increases.

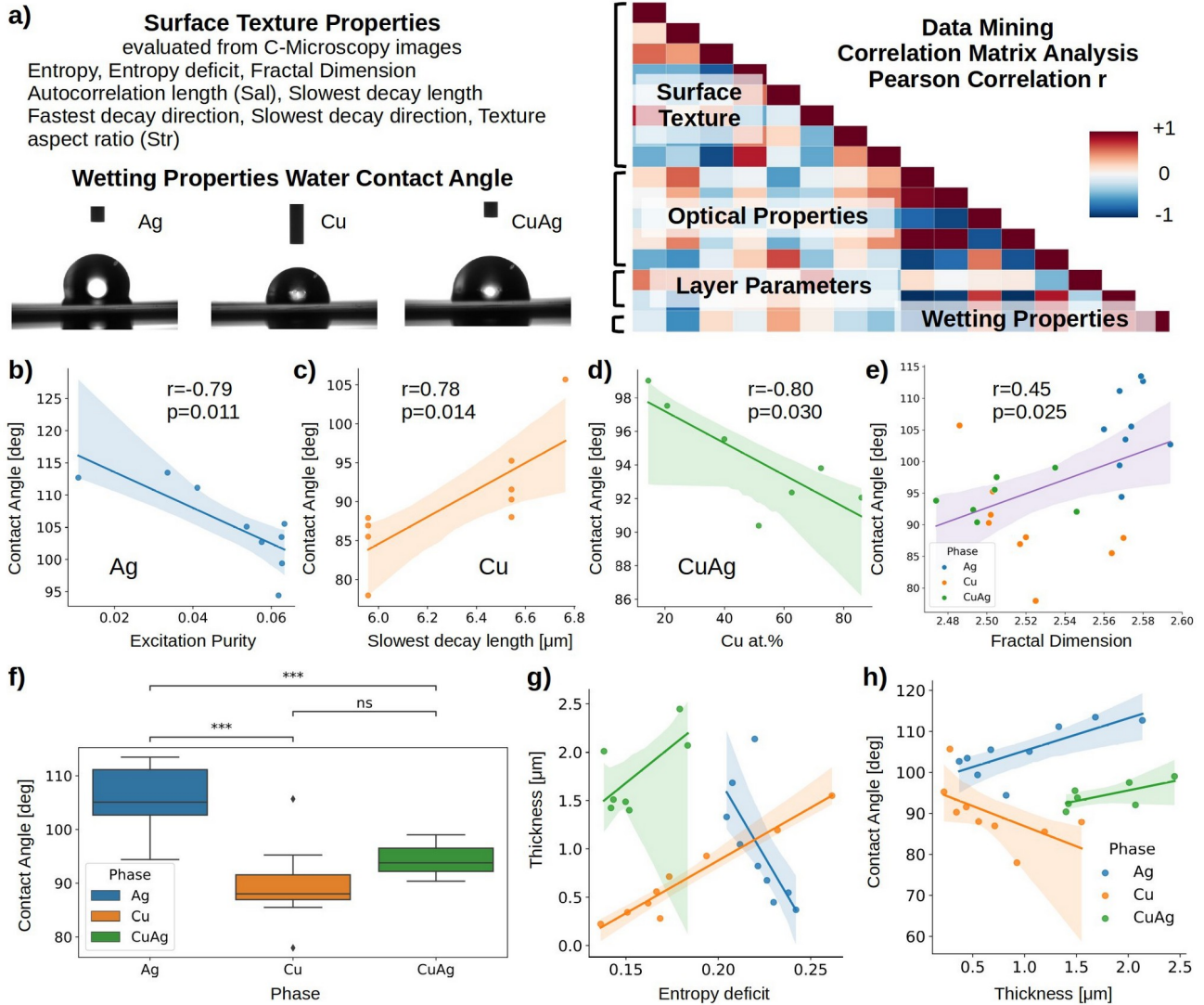

Figure 3: Characterization of CuAg samples surfaces utilizing surface texture morphology analysis, optical measurements, and wetting contact angle experiments. A color-coded correlation matrix demonstrates statistically significant relationships linking surface texture morphology properties, optical layer parameters, and wetting characteristics a). The analysis identified the following key statistically significant (p<0.05) correlations between contact angle and: Ag layer excitation purity b), the slowest decay length observed in Cu layers c), copper concentration in CuAg layers d), the fractal dimension across all layer types f). Box plot showing contact angle variation across different layers (phases) f). Discovered universal relationships: layer thickness versus entropy deficit g) and contact angle versus layer thickness h). The used correlation analysis successfully identified and discovered relations between different measured properties of CuAg samples.

This promotes the Cassie-Baxter State[33,34], where the droplet sits on top, supported by air pockets trapped in surface texture morphology. In contrast to Wenzel state[35,36], where the droplet penetrates the surface roughness and wets the surface more completely, leading to lower contact angles. This effect is important in applications like designing water-repellent coatings or optimizing surfaces for specific fluid interactions[37,38]. We also discovered universal relationships for all CuAg phases. We see that the layer thickness is correlated with a surface entropy deficit Fig. 3g), suggesting that surface order/disorder builds up during the metal growth.

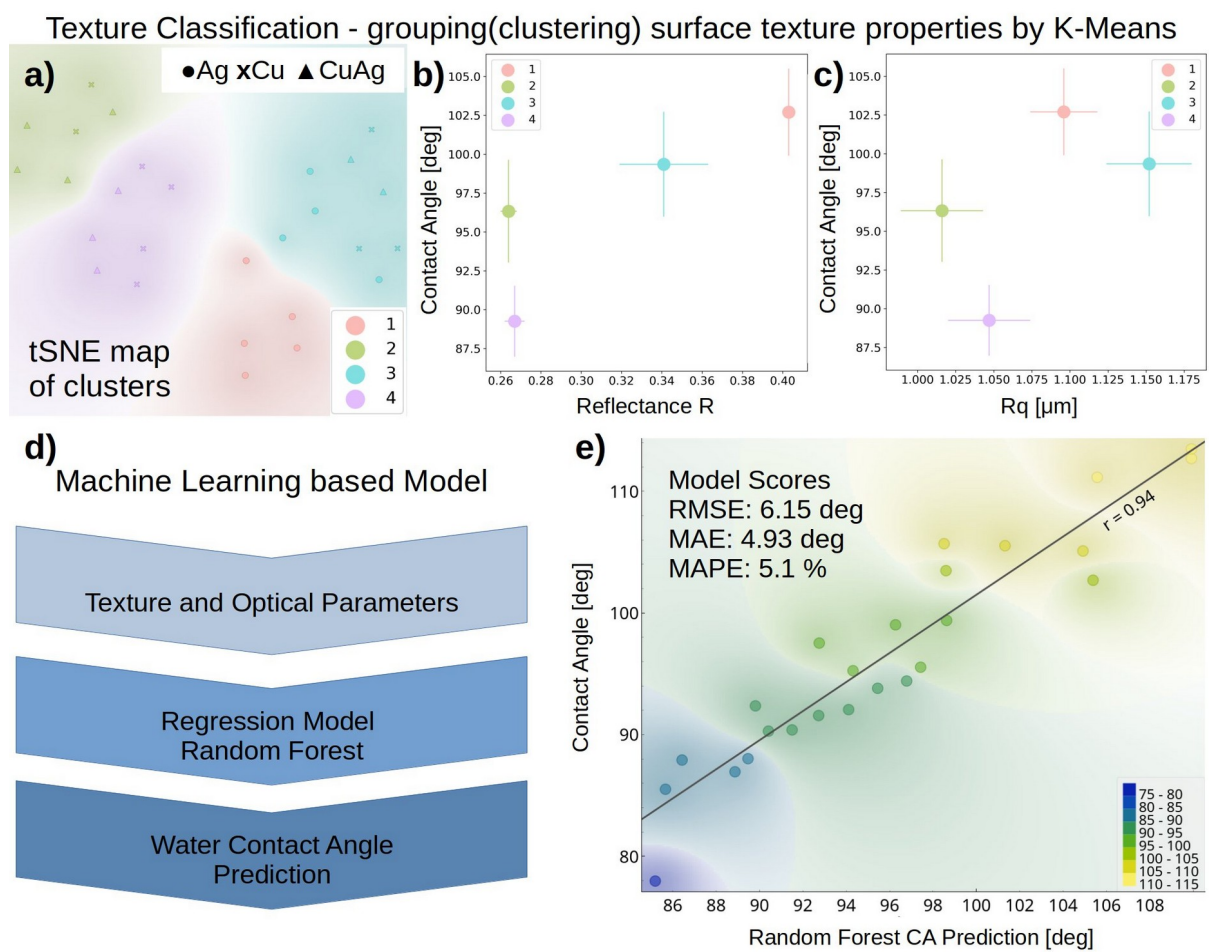

*Figure 4: CuAg surface texture morphology classification and machine learning based prediction of contact angle. a) t-Distributed Stochastic Neighbor Embedding (t-SNE) plot visualizing the distribution of four distinct surface texture morphology groups, clustered using K-Means based on similarity in surface texture morphology properties only. b) Scatter plot of contact angle versus reflectance for each identified texture group. c) Scatter plot of contact angle versus root mean square (rms) roughness for each texture group. It is seen that different surface texture morphology groups have a different roughness, wetting and optical properties. d) Diagram representing the idea of machine learning model developed for predicting water contact angle, utilizing both textural and optical properties as input. e) Comparison of measured and predicted water contact angles. The model successfully predicts the water contact angle based on the textural and optical properties.*

We also see that the contact angle depends on the layer thickness Fig. 3h). This actually suggests that the contact angle is sensitive to the complex microstructure evolution during layer growth, such as surface smoothing or roughening, which directly influences the solid area fraction $f_s$ (fraction of the solid surface in contact with the liquid) and the balance between Cassie–Baxter and Wenzel wetting regimes, see Fig. S11 in Supporting Information. For instance, as we see in our data (Fig. S11), for Cu films increasing thickness leads to enhanced smoothing, reducing air-pocket retention and shifting the surface toward Wenzel-like behavior. Conversely, in Ag films, higher thicknesses are associated with greater morphological roughening, promoting the formation and stabilization of air pockets and preserving Cassie–Baxter wetting. In the CuAg alloy, the competing effects of Cu-induced smoothing and Ag-induced roughening result in a more balanced, moderate evolution of $f_s$, leading to a relatively stable contact angle despite thickness variations. This behavior is further supported by the higher fractal dimension of the Ag phase compared to Cu and CuAg, see Fig. S12 in Supporting Information, indicating a more complex and hierarchical surface morphology. Consequently, the observed thickness dependence of wetting behavior is not merely a superficial effect, but rather a direct manifestation of the interplay between different complex effects during layer growth (grain growth, mechanical stresses, interface energy minimization) and the resulting microstructural evolution. The balance between smoothing and roughening in the CuAg alloy thus contributes to a robust and tunable wetting response, highlighting the importance of phase-specific morphological dynamics in determining surface functionality. As we showed, all key correlations and the evolution of surface texture morphology in the formed metallic layers were successfully captured through C-Microscopy measurements.

To investigate potential relationships between surface texture morphology, wetting, and optical parameters, we performed a grouping of the surface texture morphology. Utilizing eight extracted surface texture morphology parameters, we classified the texture into four distinct groups/clusters using the K-Means clustering algorithm. The number of texture groups was determined via the AIC metric, see Supporting Information Fig. S6. Fig. 4a) presents a t-Distributed Stochastic Neighbor Embedding (t-SNE) plot visualizing the distribution of these four surface texture morphology groups, projecting them from an eight-dimensional texture parameter space. Each group represents a surface texture morphology characterized by similar surface properties. Fig. 4b) displays a scatter plot of contact angle versus reflectance for each identified texture group, while Fig. 4c) shows a scatter plot of contact angle versus root mean square (rms) roughness for each texture group. These plots reveal that the different surface texture morphology groups exhibit varying roughness, wetting, and optical properties. This suggests a strong dependence between surface texture morphology characteristics and these key surface properties, potentially informing strategies for tailoring surface properties for specific applications. This also suggests that the observed differences in surface texture morphology fundamentally influence how liquids interact with and reflect light from the surface. In addition, our grouping analysis provides valuable insight into how surface texture morphology can be manipulated to achieve desired performance characteristics.

Finally, we utilized extracted texture parameters alongside optical parameters (CIE x, CIE y) to build a machine learning-based regression model for water contact angle prediction Fig. 4d). For the regression, we employed a Random Forest regression model. We assessed the model's true performance using the "leave one out" technique, assessing the model's true performance by randomly omitting one data point during training and testing. Fig. 4e) demonstrates a comparison of measured and predicted water contact angles. The model successfully predicts the water contact angle based on the textural and optical properties.

| Model Score | True Performance | Performance on training data |
| --- | --- | --- |
| RMSE (Root Mean Squared Error) | 6.154° | 3.280° |
| MAE (Mean Absolute Error) | 4.932° | 2.616° |
| MAPE (Mean Absolute Percentage Error) | 5.1% | 2.7% |

*Table 1: Evaluated Random Forest regression model performance for water contact angle prediction. The true performance was evaluated using the "leave one out" technique, assessing the model's performance by randomly omitting one data point during training and testing.*

Table 1 presents the evaluated model performance. Notably, the model achieves a mean average error (MAE) of 4.93°, which falls within the typical order of 5°, representing the typical absolute uncertainty of contact angle measurements[39]. We also conducted model explanation analysis using SHAP value analysis[40] and Permutation Feature Importance[41] to determine feature importance, see Supporting information Fig. S7-S8. Both methods identified the six most influential features driving the model's output: CIE x, CIE y, Fractal Dimension, Slowest decay length, Fastest decay direction, and Entropy deficit. This finding aligns with our correlation matrix analysis performed during the data mining stage, confirming the model's validity and transparency.

Consequently, the proposed model could be utilized to predict water contact angles for real, as-deposited metallic layers on textured surfaces under ambient conditions, potentially omitting the need for direct measurements. This also could be readily implemented in process monitoring and quality control applications. This offers a significant advantage in terms of efficiency and reduces the potential for measurement variability of the surface wetting properties. The final prediction model together with exemplary Python code is freely available in the Zenodo repository https://doi.org/10.5281/zenodo.18623766.

**Summary and Conclusions**

We explored the relationship between surface texture morphology, layer composition, wetting, and optical properties of Cu, Ag, and CuAg thin films deposited on a surface textured silicon substrate via DC magnetron sputtering. Utilizing data mining techniques and machine learning prediction, we identified key correlations - specifically, a robust link between water contact angle and surface fractal dimension across all layer types. This promotes Cassie-Baxter surface state, where a droplet sits on top of the air pockets. Furthermore, we uncovered universal relationships governing the behavior of all CuAg phases. Our analysis revealed a correlation between layer thickness and surface entropy deficit, indicating that surface order/disorder evolves during metal growth. Notably, the contact angle proved sensitive to layer thickness, suggesting an association with microstructure evolution - as thickness increases. Through K-Means clustering, we found four distinct surface texture morphology groups, each characterized by unique surface texture morphology parameters. Finally, we developed a Random Forest regression model to predict water contact angles, leveraging a combination of extracted texture features and optical parameters, achieving a mean absolute error of approximately 5°. This predictive model holds considerable promise for circumventing the need for direct measurements, on real, as-deposited metallic surfaces under ambient conditions, offering significant advantages in terms of efficiency and minimizing measurement variability. The final prediction model, alongside exemplary Python code, is freely available in the Zenodo repository. These findings transcend a purely descriptive approach, establishing a foundation for targeted surface engineering strategies focused on achieving defined wetting and optical properties through optimized surface texture morphology manipulation. By correlating texture parameters with film behavior, we've moved beyond simple characterization to a stage where precise control over surface morphology can be leveraged to tailor material performance.

**Materials and Methods**

*CuAg Film Preparation*

Films of copper (Cu) and silver (Ag), incorporating a gradient in thickness, alongside a CuAg material library, were fabricated via physical vapor deposition (PVD) using direct current magnetron sputtering (DCMS) on surface textured silicon substrates, back side of polished (100) silicon wafer. High-purity Cu and Ag targets (99.99% purity, sourced from HMW Hauner GmbH) were utilized in a co-sputtering configuration within a vacuum chamber (Korvus Technology, United Kingdom). The deposition environment was maintained at a base pressure of approximately $1 \times 10^{-6}$ mbar, utilizing argon (99.9999% purity) at a working pressure of $1 \times 10^{-2}$ mbar. The chamber was equipped with rotary and turbo pumps. Samples were divided into nine discrete, 5 mm diameter circular slices using a stainless steel mask. The substrate-target distance was meticulously controlled to achieve a pronounced thickness gradient for the Cu and Ag films, and a significant compositional gradient within the CuAg material library. Deposition powers were set at 82 W for Cu and 57 W for Ag, with a total deposition time of 1 hour per process, repeated three times on identical silicon wafers.

*Sample Characterization*

Material composition and film thickness were subsequently determined using X-ray fluorescence (XRF) spectrometry (Fischerscope X-Ray XDV-SDD, Fischer, Sindelfingen, Germany) with a 50 kV beam energy and a spot size of approximately 0.33 mm. Structural characterization, including phase identification, was performed via X-ray diffraction (XRD) using a D8 Discover diffractometer (Bruker, Billerica, USA) with CuKα1 ($\lambda = 1.5406$ Å) and CuKα2 ($\lambda = 1.54439$ Å) radiation. XRD measurements were conducted on selected regions of the highest Cu and Ag thickness, alongside all nine segments of the CuAg material library, over a 2θ range of 20-100° under conditions of 40 kV voltage, 40 mA current, and a step size of 0.02° with a collection time of 1 s per step. A -4° θ/2θ scan was implemented to minimize the contribution of the substrate's (100) reflection. Finally, surface height maps were assessed using a Sensofar 3D optical profilometer in confocal mode.

Thin electron-transparent lamellae (cross sections) for transmission Kikuchi diffraction (TKD) were prepared using a Thermo Fisher Scientific Helios 5 CXe dual-beam scanning electron microscope (SEM) equipped with a Xe plasma ion source. Prior to ion milling, a protective Pt layer was deposited by electron-beam-induced deposition (EBID) on the surface of the region of interest to minimize curtaining effects and to protect the near-surface microstructure during subsequent high-energy ion milling. Site-specific lamellae were fabricated using standard lift-out and thinning procedures.

TKD measurements were performed in the same dual-beam SEM operated at an accelerating voltage of 30 kV and a probe current of 6.4 nA, using an EBSD detector (Oxford Instruments Symmetry S3). Diffraction patterns were acquired at a resolution of 156 × 128 pixels (Speed 2 mode). The collected TKD datasets, including stored diffraction patterns, were processed using dynamic template matching implemented in Oxford Instruments AZtecCrystal MapSweeper (version 3.3 SP1)

Surface texture morphology and optical properties were measured by colorimetric microscopy (C-Microscopy)[42] utilizing digital optical microscopy and color calibration. Note that the term "texture" in this work is not used to describe the crystallographic orientation distribution of the grains. The study employed a Delta Optical Smart 5MP PRO digital optical microscope with a 5MP CMOS sensor, capturing reflected light images. A ColorChecker chart was utilized for color calibration. Recorded RGB values were converted to CIE 1931 xy chromaticity coordinates (D65 illuminant, 2deg standard observer) and later reconstructing spectral reflectance R in visible range. Using the C-Microscopy approach the color calibrated image of the CuAg samples surfaces was recorded and analyzed.

The static water contact angle experiments were performed with distilled water by the sessile drop method instrument (Attention Optical Tensiometer by Biolin Scientific) together with dedicated software OneAttention 16 (r2396) which was used to record a series of images for each measurement and later fit each one with a circle model to determined the contact angle.

The surface texture morphology properties were evaluated from C-Microscopy images using free software Gwyddion[43].

All measurements and characterizations of the Cu-Ag films were conducted on real, as-deposited surfaces under ambient conditions, including the presence of natural native oxides and surface contaminants.

The multivariate statistical correlation analysis together with K-Means clustering and Machine Learning based regression model Random Forest were performed by utilizing different free Python libraries like Scikit-learn[44], Numpy[45], Scipy[46], Orange[47], Seaborn[48].

**Acknowledgments**

This research was supported in part by the Excellence Initiative - Research University Program at the Jagiellonian University in Krakow. We would like to acknowledge Prof. Jakub Rysz and Prof. Krzysztof Dzierżega for the access to the contact angle measurements equipment. K.W. was supported by the Polish National Agency for Academic Exchange under the Polish Returns Programme.

**Author Contributions**

K.W. and J.M. contributed to the sample preparations and to the characterization by XRF, XRD and confocal microscopy. J.M. supervised the sample preparations. G.C. and P.B. contributed to the SEM transmission Kikuchi diffraction (TKD) measurements and analysis. B.R.J. conceived the idea, supervised the project and contributed to the C-Microscopy and water contact angle measurements together with multivariate and machine learning data analysis and interpretation. B.R.J. prepared the manuscript in consultation with all authors. All authors contributed to the discussion and interpretation of the final results.

**Conflicts of Interest**

The authors declare no conflicts of interest.

**Data Availability**

All the collected data, along with the developed model for water contact angle prediction based only on surface texture morphology and optical properties together with an exemplary Python program code, are freely available from the Zenodo repository https://doi.org/10.5281/zenodo.18623766.

# References

1. Zhou, X., Zhao, L., Yan, C., et al. Thermally stable threshold selector based on CuAg alloy for energy-efficient memory and neuromorphic computing applications. Nat. Commun. 14, 3285 (2023). https://doi.org/10.1038/s41467-023-39033-z
2. Benasciutti, D., Srnec Novak, J., Moro, L., & De Bona, F. Experimental characterisation of a CuAg alloy for thermo-mechanical applications. Part 1: Identifying parameters of non-linear plasticity models. Fatigue Fract. Eng. Mater. Struct., 41, 1364–1377 (2018). https://doi.org/10.1111/ffe.12783
3. Jian, C.-C., Zhang, J., & Ma, X. Cu–Ag alloy for engineering properties and applications based on the LSPR of metal nanoparticles. RSC Adv., 10(22), 13277–13285 (2020). https://doi.org/10.1039/D0RA01474E
4. Wang, D., Jung, H.D., Liu, S., et al. Revealing the structural evolution of CuAg composites during electrochemical carbon monoxide reduction. Nat. Commun. 15, 4692 (2024). https://doi.org/10.1038/s41467-024-49158-4
5. Isobayashi, A., Enomoto, Y., Yamada, H., Takahashi, S., & Kadomura, S. (2004). Thermally robust Cu interconnects with Cu-Ag alloy for sub 45nm node. (2004). IEDM Technical Digest. IEEE International Electron Devices Meeting, San Francisco, CA, USA, 2004, pp. 953–956. http://doi.org/10.1109/IEDM.2004.1419342.
6. Strehle, S., Menzel, S., Bartha, J.W., & Wetzig, K. (2010). Electroplating of Cu(Ag) thin films for interconnect applications. Microelectronic Engineering, 87, 180–186. https://doi.org/10.1016/j.mee.2009.07.010.
7. Kong, L.W., Zhu, X.L., Xing, Z.B., Chang, Y.Q., Huang, H., Shu, Y., Qi, Z.X., & Wen, B. (2024). Preparation and mechanisms of Cu–Ag alloy fibers with high strength and high conductivity. Materials Science and Engineering: A, 895, 146219. https://doi.org/10.1016/j.msea.2024.146219
8. Park, J., Ahn, M., Yu, G., Kim, J., & Kim, S. (2024). Influence of alloying elements and composition on microstructure and mechanical properties of Cu-Si, Cu-Ag, Cu-Ti, and Cu-Zr alloys. Materials Today Communications, 38, 107821. https://doi.org/10.1016/j.mtcomm.2023.107821
9. Bernasconi, R., Hart, J.L., Lang, A.C., Magagnin, L., & Nobili, L. (2017). Structural properties of electrodeposited Cu-Ag alloys. Electrochimica Acta, 251, 475–481. https://doi.org/10.1016/j.electacta.2017.08.097
10. Wieczerzak, K., Groetsch, A., Pajor, K., Jain, M., Müller, A.M., Vockenhuber, C., Schwiedrzik, J., Sharma, A., Klimashin, F.F., & Michler, J. (2023). Unlocking the Potential of CuAgZr Metallic Glasses: A Comprehensive Exploration with Combinatorial Synthesis, High-Throughput Characterization, and Machine Learning. Adv. Sci., 10, 2302997. https://doi.org/10.1002/advs.202302997
11. Pajor, K., Jain, M., Kozieł, T., Pikulski, D., Bała, P., Michler, J., & Wieczerzak, K. (2025). Comparative analysis of plastic deformation in Zr-Cu-Ag metallic glasses across macro and micro scales: Insights from micropillar and bulk sample compression tests of fully and partially amorphous alloy. Journal of Alloys and Compounds, 1031, 181053. https://doi.org/10.1016/j.jallcom.2025.181053
12. Wu, Z.L., Buguin, A., Yang, H., Taulemesse, J.-M., Le Moigne, N., Bergeret, A., Wang, X., & Keller, P. (2013). Microstructured nematic liquid crystalline elastomer surfaces with switchable wetting properties. Adv. Func. Mater., 23, 3070–3076. https://doi.org/10.1002/adfm.201203291
13. Ge, P., Wang, S., Zhang, J., & Yang, B. (2020). Micro-/nanostructures meet anisotropic wetting: from preparation methods to applications. Mater. Horiz., 7(10), 2566–2595. https://doi.org/10.1039/D0MH00768D
14. Zhang, X., Scaraggi, M., Zheng, Y., Li, X., Wu, Y., Wang, D., Dini, D., & Zhou, F. (2022). Quantifying wetting dynamics with triboelectrification. Adv. Sci., 9(22), 2200822. https://doi.org/10.1002/advs.202200822
15. Shen, Z., Wu, H., Liu, C., et al. (2026). Wafer-scale monolayer dielectric integration on atomically thin semiconductors. Nature Materials, 25, 1–10. https://doi.org/10.1038/s41563-025-02445-x
16. Li, Y., Sasaki, T., Shimizu, Y., & Koshizaki, N. (2008). A hierarchically ordered $TiO_2$ hemispherical particle array with hexagonal-non-close-packed tops: Synthesis and stable superhydrophilicity without UV irradiation. Small, 4(8), 2286–2291. https://doi.org/10.1002/smll.200800428
17. Wang, F., Wang, C., Wei, D., Li, G., Zhang, W., & Zhao, Z. (2025). Engineering thin water film and cluster evaporation towards extraordinarily high 2D solar vapor generation. Materials Today, 90, 258–269. https://doi.org/10.1016/j.mattod.2025.09.023
18. Min, L., Xiaojie, S., Peipei, L., & Meiping, W. (2022). Forming quality and wettability of surface texture morphology on CuSn10 fabricated by laser powder bed fusion. AIP Advances, 12(12), 125114. https://doi.org/10.1063/5.0122076
19. Yonehara, M., Matsui, T., Kihara, K., Isono, H., Kijima, A., & Sugibayashi, T. (2005). Experimental relationships between surface roughness, glossiness and color of chromatic colored metals. Materials Transactions, 45(4), 1027–1032. https://doi.org/10.2320/matertrans.45.1027
20. Vishnoi, M., Kumar, P., & Murtaza, Q. (2021). Surface texturing techniques to enhance tribological performance: A review. Surfaces and Interfaces, 27, 101463. https://doi.org/10.1016/j.surfin.2021.101463
21. Hsieh, J., & Hung, S. The Effect of Cu:Ag Atomic Ratio on the Properties of Sputtered Cu–Ag Alloy Thin Films. Materials 2016, 9, 914. https://doi.org/10.3390/ma9110914
22. Jany, B. R., Janas, A., & Krok, F. (2017). Retrieving the Quantitative Chemical Information at Nanoscale from Scanning Electron Microscope Energy Dispersive X-ray Measurements by Machine Learning. Nano Letters, 17(11), 6507–6508. https://doi.org/10.1021/acs.nanolett.7b01789

# References

23  Jany, B. R., Janas, A., & Krok, F. (2020). Automatic microscopic image analysis by moving window local Fourier Transform and Machine Learning. Micron, 130, 102800. https://doi.org/10.1016/j.micron.2019.102800

24  Jany, B. R. (2025). Machine Learning techniques in microscopic characterisation of nanomaterials. In OP Conference Series on Materials Science and Engineering, 1324, 012007. https://doi.org/10.1088/1757-899X/1324/1/012007

25  Jany, B. R. (2026). Collected Experimental Data For "Exploring Wetting and Optical Properties of CuAg Alloys via Surface Texture Morphology Analysis" [Data set]. Zenodo. https://doi.org/10.5281/zenodo.18623766

26  Elliott, R.P., Shunk, F.A., & Giessen, W.C. (1980). The Ag−Cu (Silver-Copper) system. Bulletin of Alloy Phase Diagrams, 1(41–45). https://doi.org/10.1007/BF02883284

27  Dulmaa, A., Cougnon, F. G., Dedoncker, R., & Depla, D. (2021). On the grain size-thickness correlation for thin films. Acta Materialia, 212, 116896. https://doi.org/10.1016/j.actamat.2021.116896

28  Gaskey, B., Hendl, L., Wang, X., & Seita, M. Optical characterization of grain orientation in crystalline materials. Acta Materialia, 194, 558–564 (2020). https://doi.org/10.1016/j.actamat.2020.05.027

29  Lozanova, V., Tasseva, J., & Todorov, R.*. (2013). Grain size effect on the optical properties of thin silver films. Bulgarian Chemical Communications, 45(Special Issue B), 43–46. https://bcc.bas.bg/BCC_Volumes/Volume_45_Special_B_2013/Volume_45_Special_B_2013_PDF/7-Vlozanova.pdf

30  Jiang, Y., Pillai, S., & Green, M. A. (2016). Grain boundary effects on the optical constants and Drude relaxation times of silver films. Journal of Applied Physics, 120(12), 123109. https://doi.org/10.1063/1.4972471

31  Somlyai-Sipos, L., & Baumli, P. (2022). Wettability of Metals by Water. Metals, 12(8), 1274. https://doi.org/10.3390/met12081274

32  Bryk, P., Korczeniewski, E., Szymański, G. S., Kowalczyk, P., Terpiłowski, K., & Terzyk, A. P. (2020). What Is the Value of Water Contact Angle on Silicon? Materials, 13(7), 1554. https://doi.org/10.3390/ma13071554

33  Murakami, D., Jinnai, H., & Takahara, A. (2014). Wetting transition from the Cassie–Baxter state to the Wenzel state on textured polymer surfaces. Langmuir, 30(8), 2061–2067. https://doi.org/10.1021/la4049067

34  Rohrs, C., Azimi, A., & He, P. Wetting on micropatterned surfaces: Partial penetration in the Cassie state and Wenzel deviation theoretically explained. Langmuir, 35(47), 15421–15430 (2019). https://doi.org/10.1021/acs.langmuir.9b03002

35  Liu, Z., Pan, N., Tao, H., et al. Temperature-dependent wetting characteristics of micro–nano-structured metal surface formed by femtosecond laser. J. Mater. Sci., 56, 3525–3534 (2021). https://doi.org/10.1007/s10853-020-05457-x

36  Park, W., Ribe, J. M., Fernandino, M., & Dorao, C. A. The criterion of the Cassie–Baxter and Wenzel wetting modes and the effect of elastic substrates on it. Adv. Mater. Interfaces, 10, 2202439 (2023). https://doi.org/10.1002/admi.202202439

37  Zhang, L. et al., (2023). Fractal theory and dynamic contact angle-based imbibition model for two-phase flow in porous media. Physics of Fluids, 35(12), 122019. https://doi.org/10.1063/5.0181498

38  Jasrotia, P., Priya, B., Kumar, R., Bishnoi, P., Vij, A., & Kumar, T. (2023). A correlation between fractal growth, water contact angle, and SERS intensity of R6G on ion beam nanostructured ultra-thin gold (Au) films. Frontiers in Physics, 11, 1125004. https://doi.org/10.3389/fphy.2023.1125004

39  Extrand, C. W. (2015). Uncertainty in contact angle estimates from a Wilhelmy tensiometer. Journal of Adhesion Science and Technology, 29(23), 2515–2520. https://doi.org/10.1080/01694243.2015.1072775

40  Lundberg, S. M., & Lee, S. I. (2017). A unified approach to interpreting model predictions. In Advances in Neural Information Processing Systems (pp. 1-11). Curran Associates, Inc. https://proceedings.neurips.cc/paper_files/paper/2017/file/8a20a8621978632d76c43dfd28b67767-Paper.pdf

41  Breiman, L. (2001). Random Forests. Machine Learning, 45(1), 5–32. https://doi.org/10.1023/A:1010933404324

42  Jany, B.R. (2024). Quantifying colors at micrometer scale by colorimetric microscopy (C-Microscopy) approach. Micron, 176, 103557. ISSN 0968-4328. https://doi.org/10.1016/j.micron.2023.103557

43  Nečas, D., & Klapetek, P. (2012). Gwyddion: an open-source software for {SPM} data analysis. Central European Journal of Physics, 10(1), 181–188. https://doi.org/10.2478/s11534-011-0096-2

44  Pedregosa, F. et al., (2011). Scikit-learn: Machine Learning in Python. Journal of Machine Learning Research, 12, 2825–2830.

45  Harris, C.R. et al., (2020). Array programming with NumPy. Nature, 585(7825), 357–362. https://doi.org/10.1038/s41586-020-2649-2

46  Virtanen, P. et al., (2020). {SciPy} 1.0: Fundamental algorithms for scientific computing in Python. Nature Methods, 17(1), 261–272. https://doi.org/10.1038/s41592-019-0686-2

47  Demsar, J. et al., (2013). Orange: Data Mining Toolbox in Python. Journal of Machine Learning Research, 14, 2349–2353.

48  Waskom, M. L. (2021). Seaborn: statistical data visualization. Journal of Open Source Software, 6(60), 3021. https://doi.org/10.21105/joss.03021

# SUPPORTING INFORMATION
# Exploring Wetting and Optical Properties of CuAg Alloys
# via Surface Texture Morphology Analysis

Krzysztof Wieczerzak[a], Grzegorz Cios[b], Piotr Bała[b,c], Johann Michler[d], Benedykt R. Jany[e*]

[a]Department of Materials Science, Faculty of Mechanical Engineering and Aeronautics, Rzeszow University of Technology, al. Powstańców Warszawy 12, 35-959 Rzeszow, Poland
[b]Academic Centre for Materials and Nanotechnology, AGH University of Krakow, al. A Mickiewicza 30, 30-059 Krakow, Poland
[c]Faculty of Metals Engineering and Industrial Computer Science, AGH University of Krakow, al. A Mickiewicza 30, 30-059 Krakow, Poland
[d]Laboratory for Mechanics of Materials and Nanostructures, Empa, Swiss Federal Laboratories for Materials Science and Technology, CH-3602 Thun, Switzerland
[e]Marian Smoluchowski Institute of Physics, Faculty of Physics, Astronomy and Applied Computer Science, Jagiellonian University, PL-30348 Krakow, Poland

## Surface Texture Morphology – Colorimetric Microscopy (C-Microscopy)

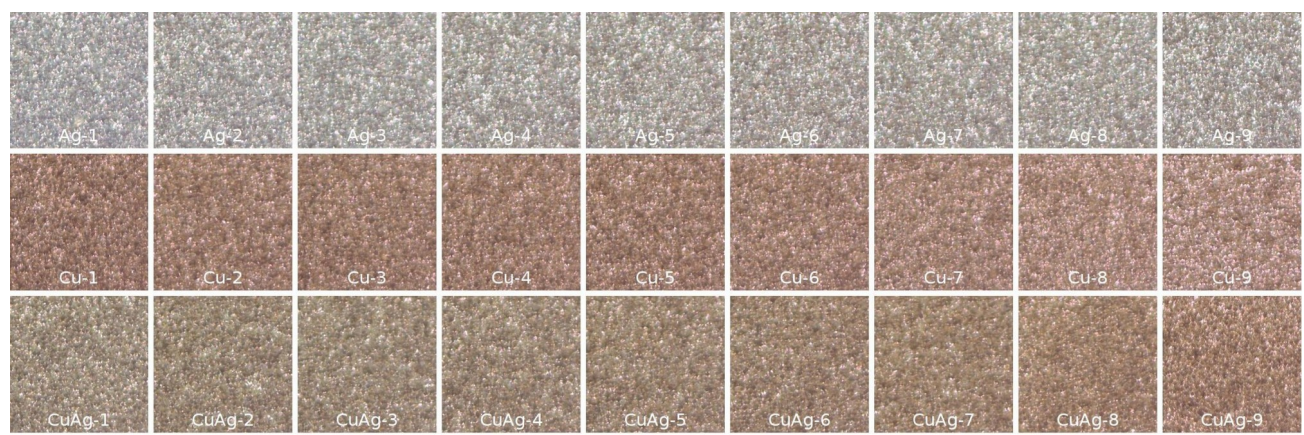

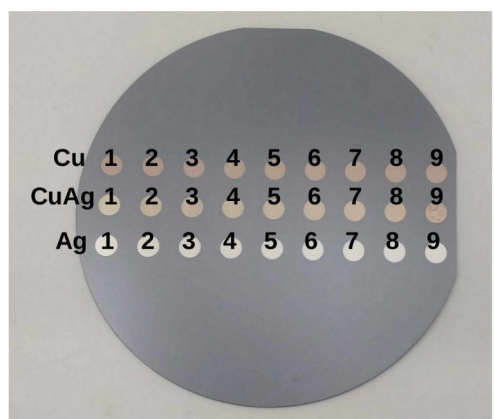

*Figure S1: Color Calibrated Images of CuAg Surfaces (D65 illuminant, CIE1931 standard observer) as measured by C-Microscopy. Upper row Ag surfaces, middle row Cu surfaces, lower row CuAg surfaces. The image width corresponds to the 438 microns. Below the scheme of CuAg Material Library as synthesized on Si. Sample number (Patch) indicated.*

---

\*   Corresponding author e-mail: benedykt.jany@uj.edu.pl

| Phase | sample number (Patch) | Thickness [µm] | Cu a.t% |
|---|---|---|---|
| Ag | 1 | 2.13667 | 0 |
| Ag | 2 | 1.68333 | 0 |
| Ag | 3 | 1.33 | 0 |
| Ag | 4 | 1.04667 | 0 |
| Ag | 5 | 0.82233 | 0 |
| Ag | 6 | 0.67433 | 0 |
| Ag | 7 | 0.547 | 0 |
| Ag | 8 | 0.44733 | 0 |
| Ag | 9 | 0.36967 | 0 |
| Cu | 1 | 0.22267 | 100 |
| Cu | 2 | 0.28 | 100 |
| Cu | 3 | 0.34333 | 100 |
| Cu | 4 | 0.438 | 100 |
| Cu | 5 | 0.55767 | 100 |
| Cu | 6 | 0.71267 | 100 |
| Cu | 7 | 0.92633 | 100 |
| Cu | 8 | 1.19333 | 100 |
| Cu | 9 | 1.55 | 100 |
| CuAg | 1 | 2.44667 | 14.4 |
| CuAg | 2 | 2.01 | 20.73 |
| CuAg | 3 | 1.70667 | 29.00 |
| CuAg | 4 | 1.48667 | 39.93 |
| CuAg | 5 | 1.4 | 51.46 |
| CuAg | 6 | 1.42333 | 62.56 |
| CuAg | 7 | 1.51 | 72.36 |
| CuAg | 8 | 1.7 | 80.3 |
| CuAg | 9 | 2.07 | 85.93 |

*Table S1: Parameters of CuAg Material Library as determined after synthesis from XRF measurements.*

# Data Mining – Pearson Correlation Matrix Analysis

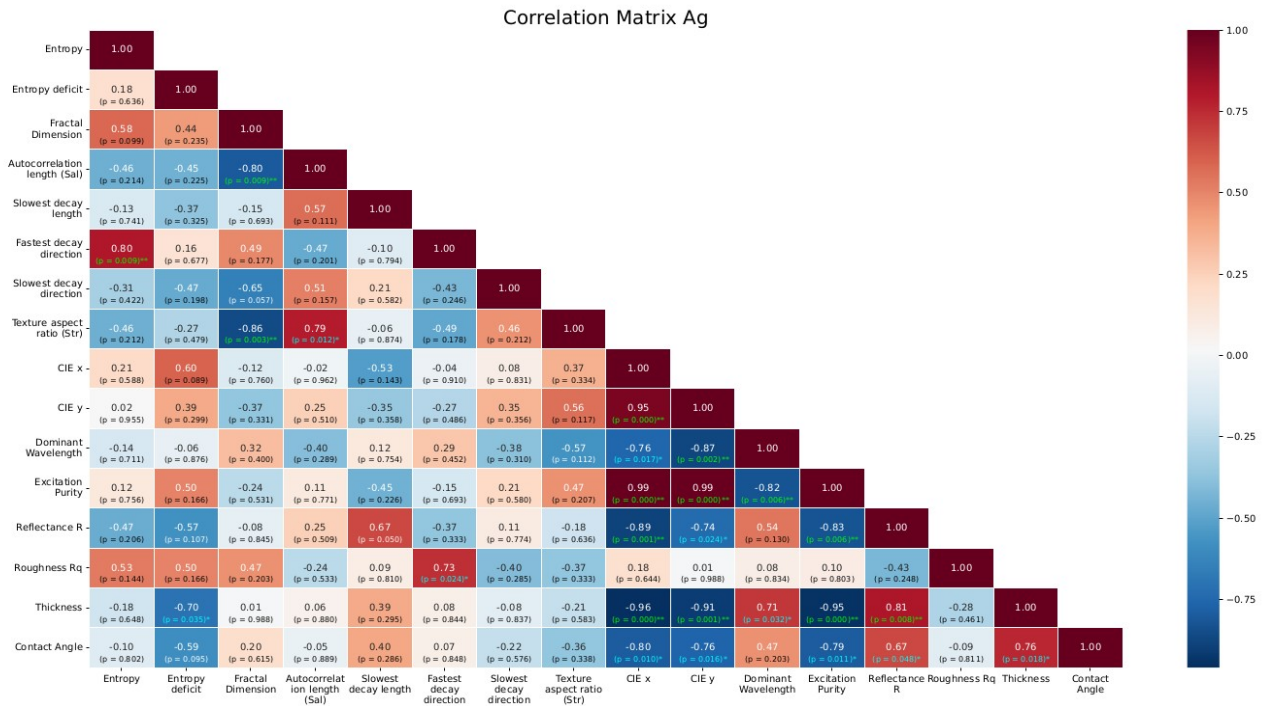

*Figure S2: Correlation matrix for the Ag phase. Statistically significant correlations marked (p<0.05).*

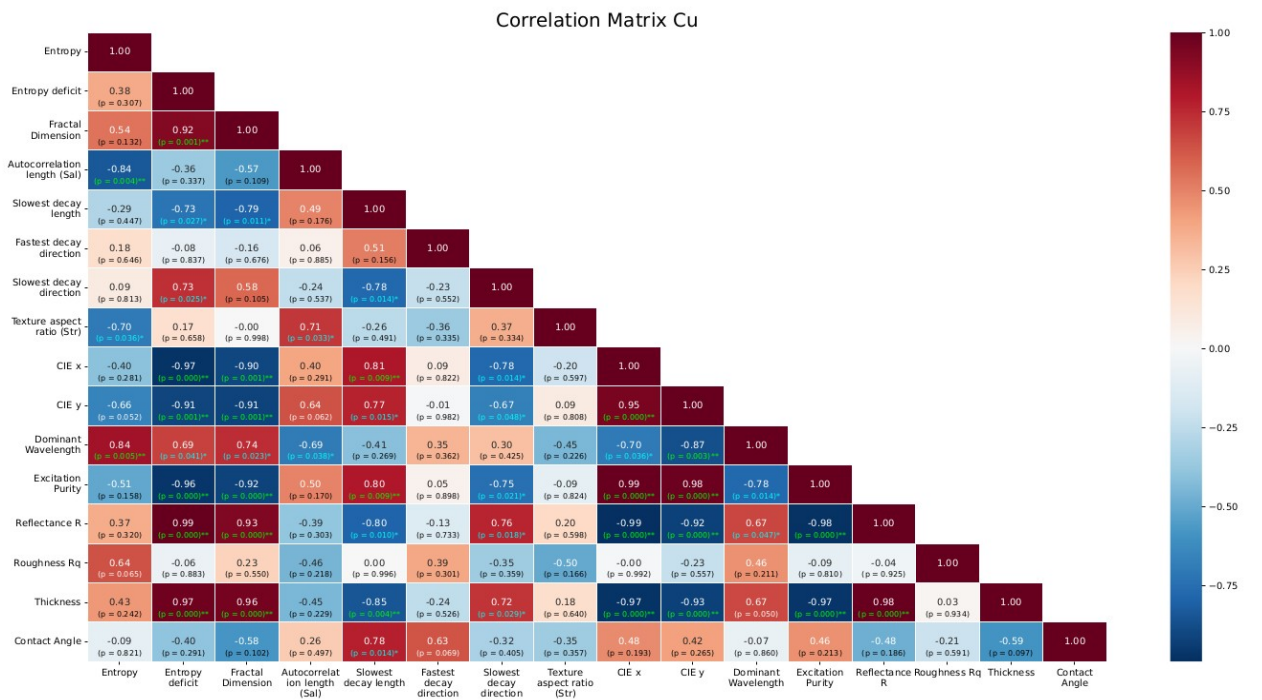

*Figure S3: Correlation matrix for the Cu phase. Statistically significant correlations marked (p<0.05).*

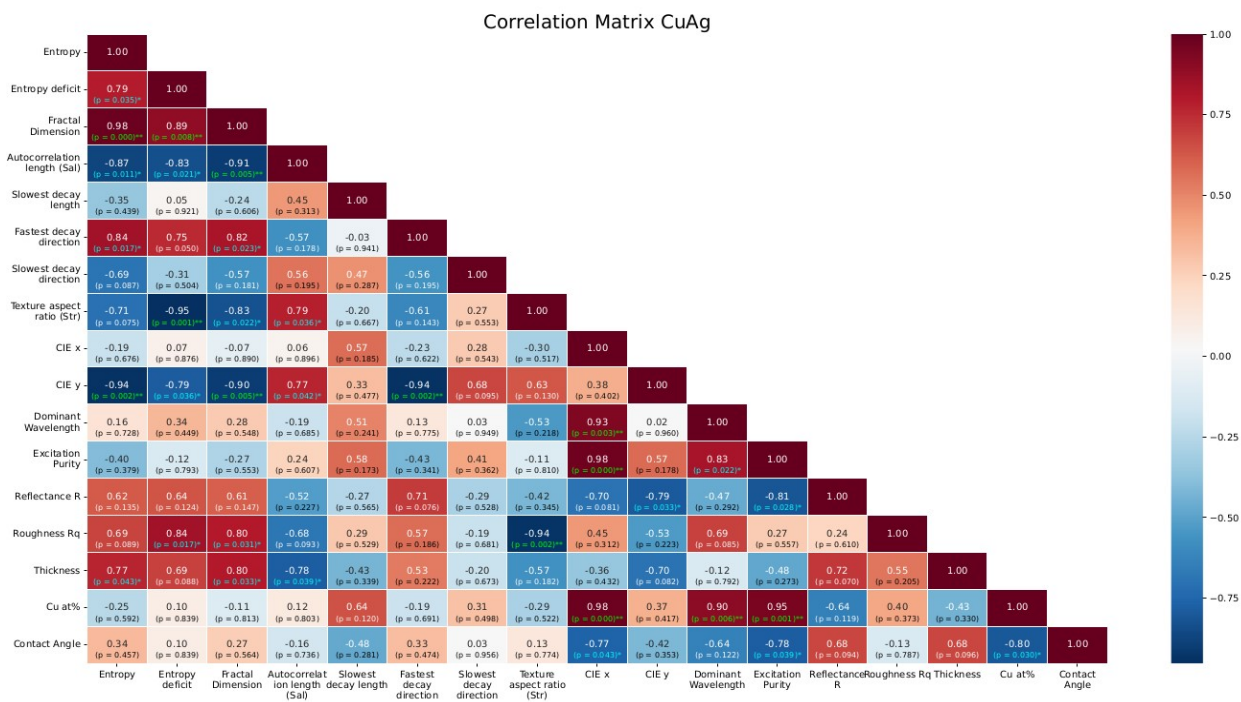

*Figure S4: Correlation matrix for the CuAg phase. Statistically significant correlations marked (p<0.05).*

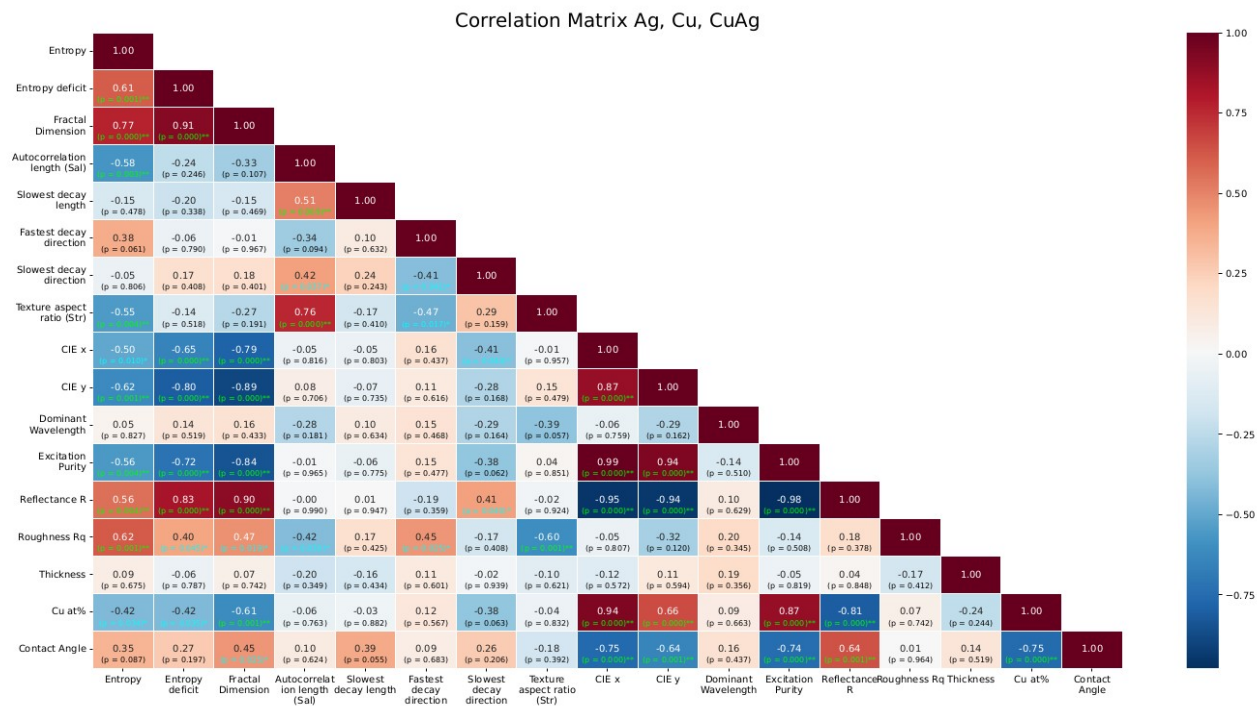

*Figure S5: Correlation matrix for the all phases together Ag, Cu and CuAg phase. Statistically significant correlations marked (p<0.05).*

**Surface Texture Morphology Classification**

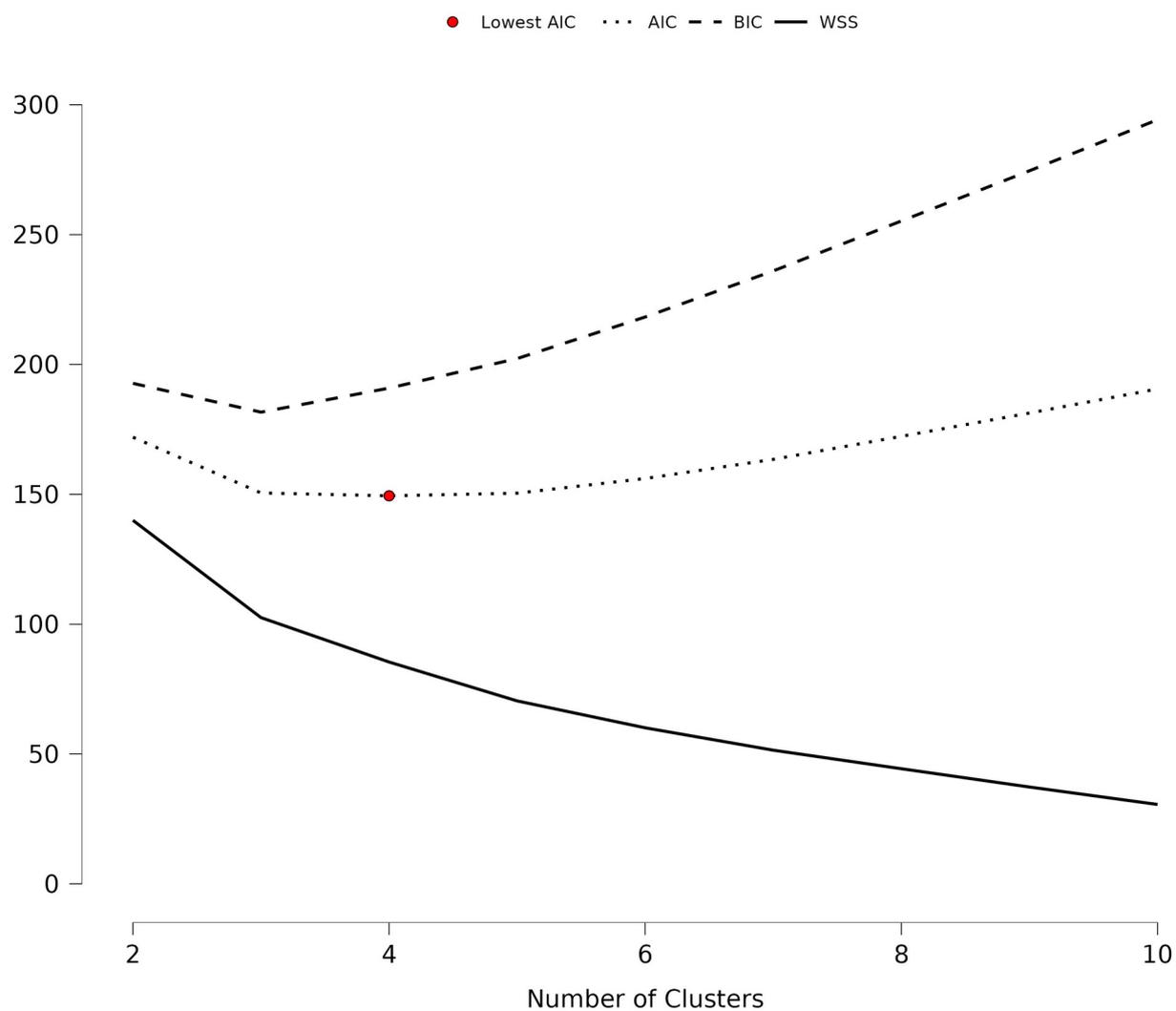

*Figure S6: Number of clusters/groups selection for surface texture morphology classification based on the information criteria. The optimal number of clusters for surface texture morphology classification was determined using the Akaike Information Criterion (AIC), selecting the number of clusters with the lowest value.*

**Explanation of Machine Learning Model for Water Contact Angle Prediction**

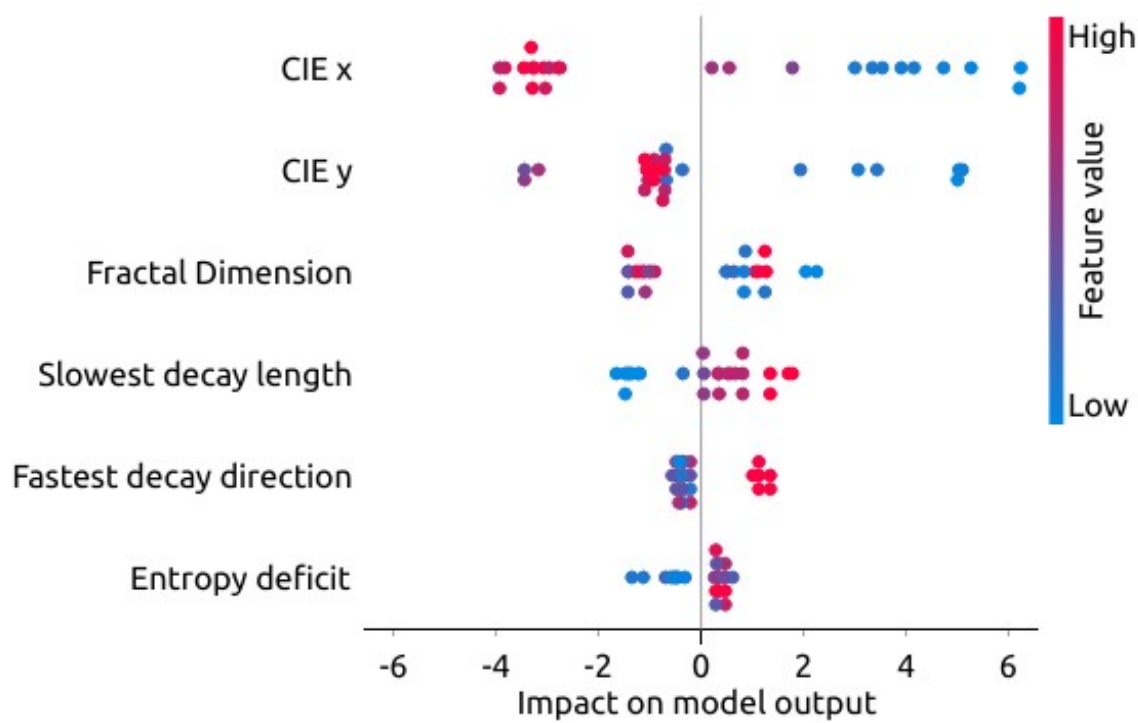

*Figure S7: Model explanation by SHAP value analysis of a Random Forest model, utilizing the SHAP library. The x-axis represents SHAP values for each data point, indicating how much each feature's value contributes to the model's prediction. A SHAP value quantifies this contribution by measuring the difference between the predicted value and the average predicted value across the dataset. The color scale represents the feature values; red denotes higher values, and blue represents lower values. Importantly, features are sorted on the y-axis according to their impact on the model's prediction, allowing us to identify the six most influential features driving the model's output i.e. CIE x, CIE y, Fractal Dimension, Slowest decay length, Fastest decay direction, Entropy deficit. It is not surprising since these features were also identified in Data Mining step – correlation matrix analysis.*

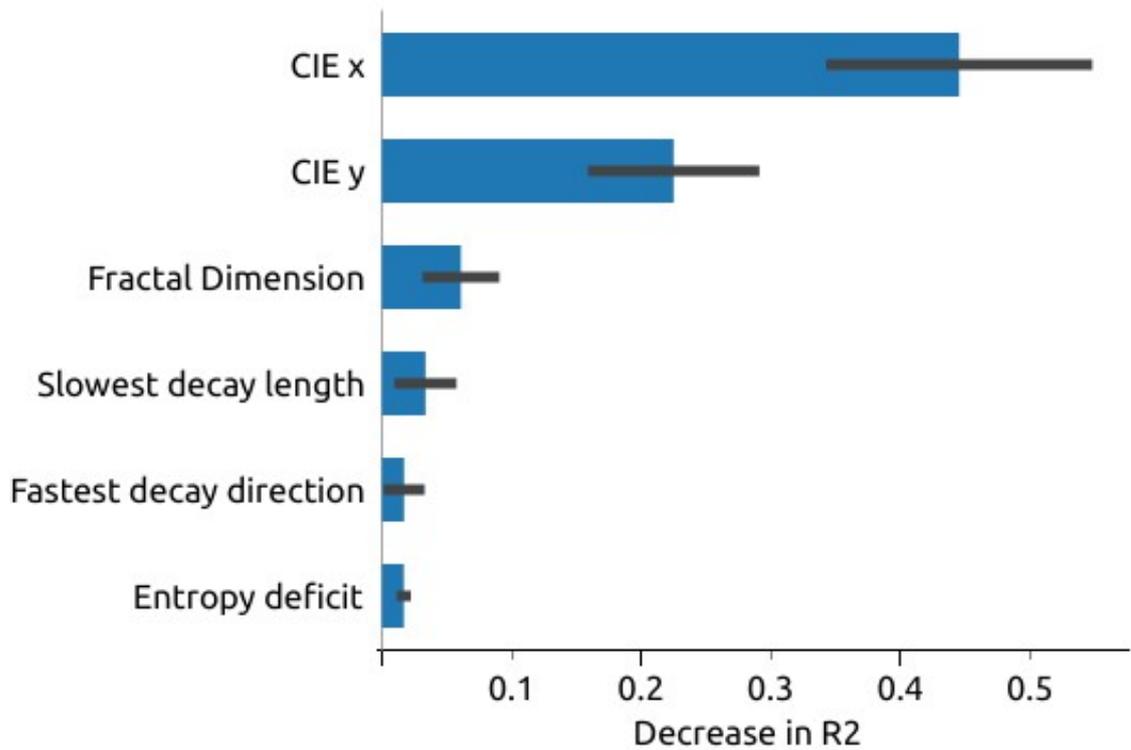

*Figure S8: Feature importance derived from a Random Forest model using Permutation Feature Importance. This technique measures each feature's contribution to prediction accuracy by calculating the increase in model prediction error following the random shuffling of the feature's values. The plot displays features ordered by their impact on the prediction error; the six features with the greatest negative impact (indicating highest importance) are shown. The following important features were identified: CIE x, CIE y, Fractal Dimension, Slowest decay length, Fastest decay direction, Entropy deficit. It is not surprising since these features were also identified by Data Mining – correlation matrix analysis.*

**Results of SEM transmission Kikuchi diffraction (TKD) measurements of CuAg alloy samples**

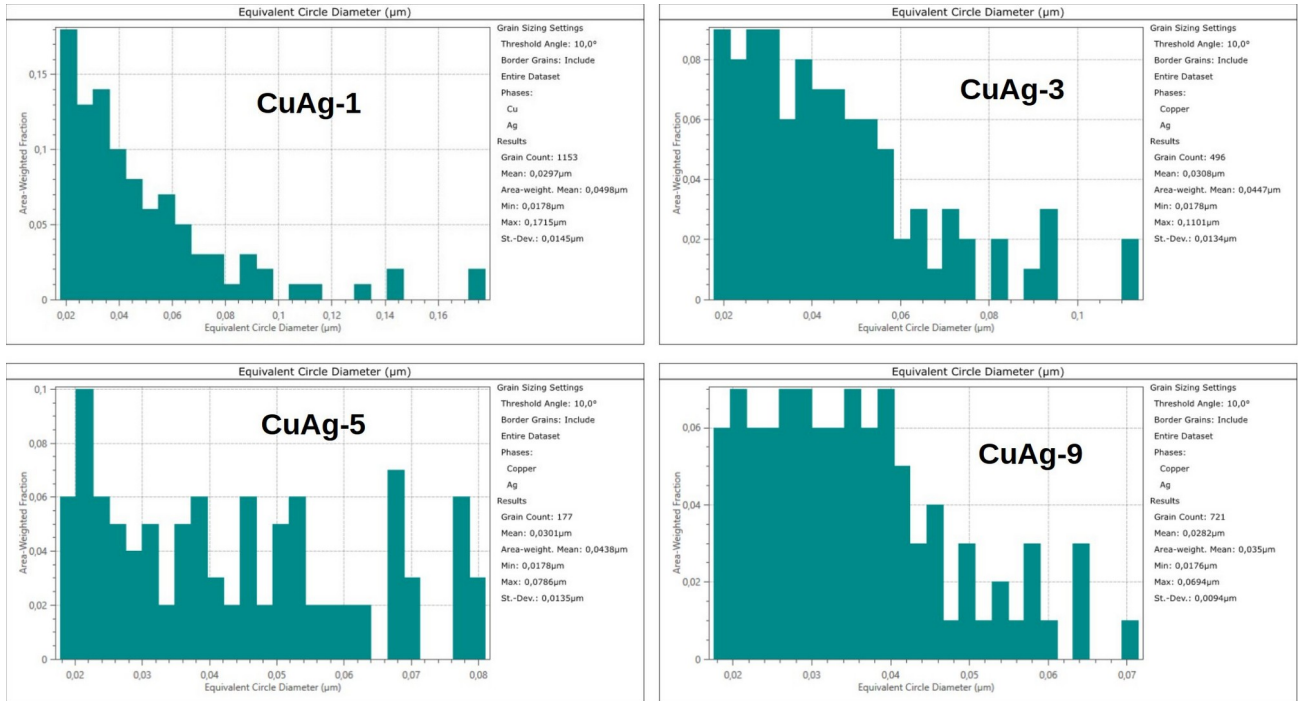

*Figure S9: Crystallographic grain distributions of CuAg-1, CuAg-3, CuAg-5, and CuAg-9 alloys determined by SEM transmission Kikuchi diffraction (TKD) measurements analysis. Grain sizes are calculated as equivalent circle diameters, and the distributions are presented as area-weighted fractions*

## Calculated Water Contact Angle on CuAg alloys flat surface

We calculated water contact angle (CA) on CuAg alloys flat surface from average atomic radius ($r_A$) of CuAg using the following empirical formula for metals:

$$CA = -0.582 * r_A + 148$$

We calculated the average atomic radius of CuAg by weighted atomic percent of the copper atomic radius (128pm) and silver atomic radius (144pm). The calculated water contact angle is for the CuAg alloy for different copper content for the flat polished surface. The formula is taken from Somlyai-Sipos, L., & Baumli, P. (2022). Wettability of Metals by Water. Metals, 12(8), 1274. https://doi.org/10.3390/met12081274. It is seen (Table S2 and Fig. S10) that as the copper content increases the water contact angle also increases.

| Phase | Cu At.% | Average Atomic Radius [pm] | Calculated Water Contact Angle [deg] |
|---|---|---|---|
| CuAg-1 | 14.40 | 141.70 | 65.53 |
| CuAg-2 | 20.73 | 140.68 | 66.12 |
| CuAg-3 | 29.00 | 139.36 | 66.89 |
| CuAg-4 | 39.93 | 137.61 | 67.91 |
| CuAg-5 | 51.46 | 135.77 | 68.98 |
| CuAg-6 | 62.56 | 133.99 | 70.02 |
| CuAg-7 | 72.36 | 132.42 | 70.93 |
| CuAg-8 | 80.30 | 131.15 | 71.67 |
| CuAg-9 | 85.93 | 130.25 | 72.19 |

*Table S2: Calculated water contact angle for the CuAg alloys for different copper content for the flat polished surface. As the copper content increases the water contact angle also increases.*

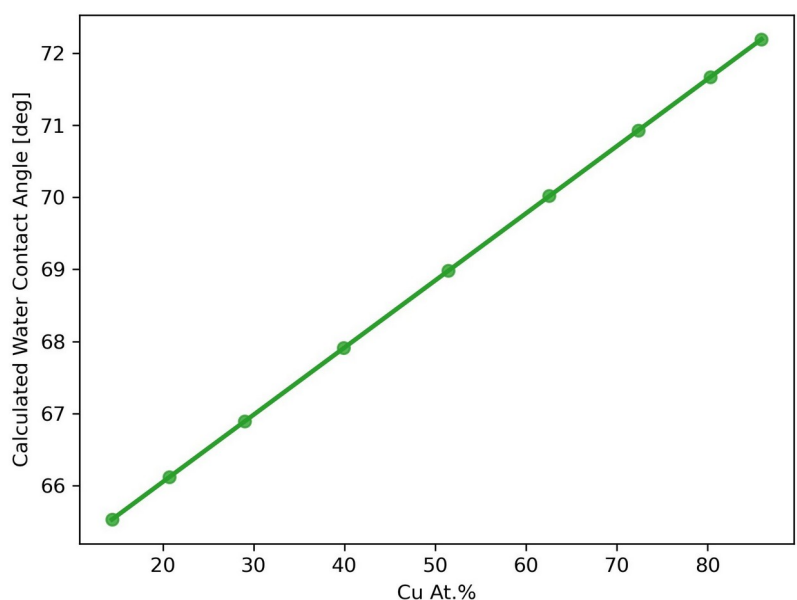

*Figure S10: Calculated water contact angle for the CuAg alloys for different copper content for the flat polished surface.*

## Cassie-Baxter Wetting State

Assuming Cassie-Baxter wetting state for the composite surface (solid+air pockets) one can relate the contact angle on the textured surface $CA_{textured}$ with the contact angle on the flat surface $CA_{flat}$ in the following way:

$$\cos(CA_{textured}) = f_s(\cos(CA_{flat})+1) - 1$$

where $f_s$ is solid are fraction i.e. fraction of the solid surface in contact with the liquid, the rest $(1-f_s)$ is the air pockets contribution, this could be expressed as:

$$f_s = \frac{\cos(CA_{textured})+1}{\cos(CA_{flat})+1}$$

Based on this we calculated the $f_s$ for our contact angle measurements for all the layer phases. For the the contact angle on the flat surface we used the values CA(Ag) = 64°, CA(Cu) = 72° and for the CuAg from the Table S2 using values and formulas from Somlyai-Sipos, L., & Baumli, P. (2022). Wettability of Metals by Water. Metals, 12(8), 1274. https://doi.org/10.3390/met12081274. The average solid area fraction $f_s$ is different for different phases $f_s$(Ag) = 0.51, $f_s$(Cu) = 0.77, $f_s$(CuAg) = 0.68, see Fig. S11a), for the comparison for our textured silicone $f_s$(Si) = 0.84.

We see that as the Cu content increases, the Cassie–Baxter solid fraction $f_s$ rises from ~0.6 to ~0.8, see Fig. S11b). This corresponds to a reduction of the trapped air fraction from ~40% to ~20%, i.e., ~50% decrease in air pocket contribution. Consequently, the surface becomes less in the Cassie–Baxter regime, leading to reduced apparent contact angle and increased droplet adhesion. The surface shifts toward more Wenzel like wetting behavior.

We see also that Solid Area Fraction $f_s$ changes with layer thickness for all the phase, see Fig. S11c). For the Cu phase the $f_s$ increases with layer thickness from ~0.7 to ~0.9, indicating progressive suppression of air pockets as the Cu layer grows. Thicker Cu layers therefore promote a more Wenzel like wetting regime. For Ag, the solid fraction $f_s$ decreases from ~0.6 to ~0.4 as layer thickness increases, showing that the Ag layer develops more pronounced air retaining microstructures. This enhances Cassie–Baxter wetting for thicker Ag films. For the CuAg alloy, the solid fraction $f_s$ decreases only slightly (from ~0.7 to ~0.6) with increasing thickness, suggesting that the competing morphological tendencies of Cu (smoothing) and Ag (roughening) lead to a moderate enhancement of air-pocket formation.

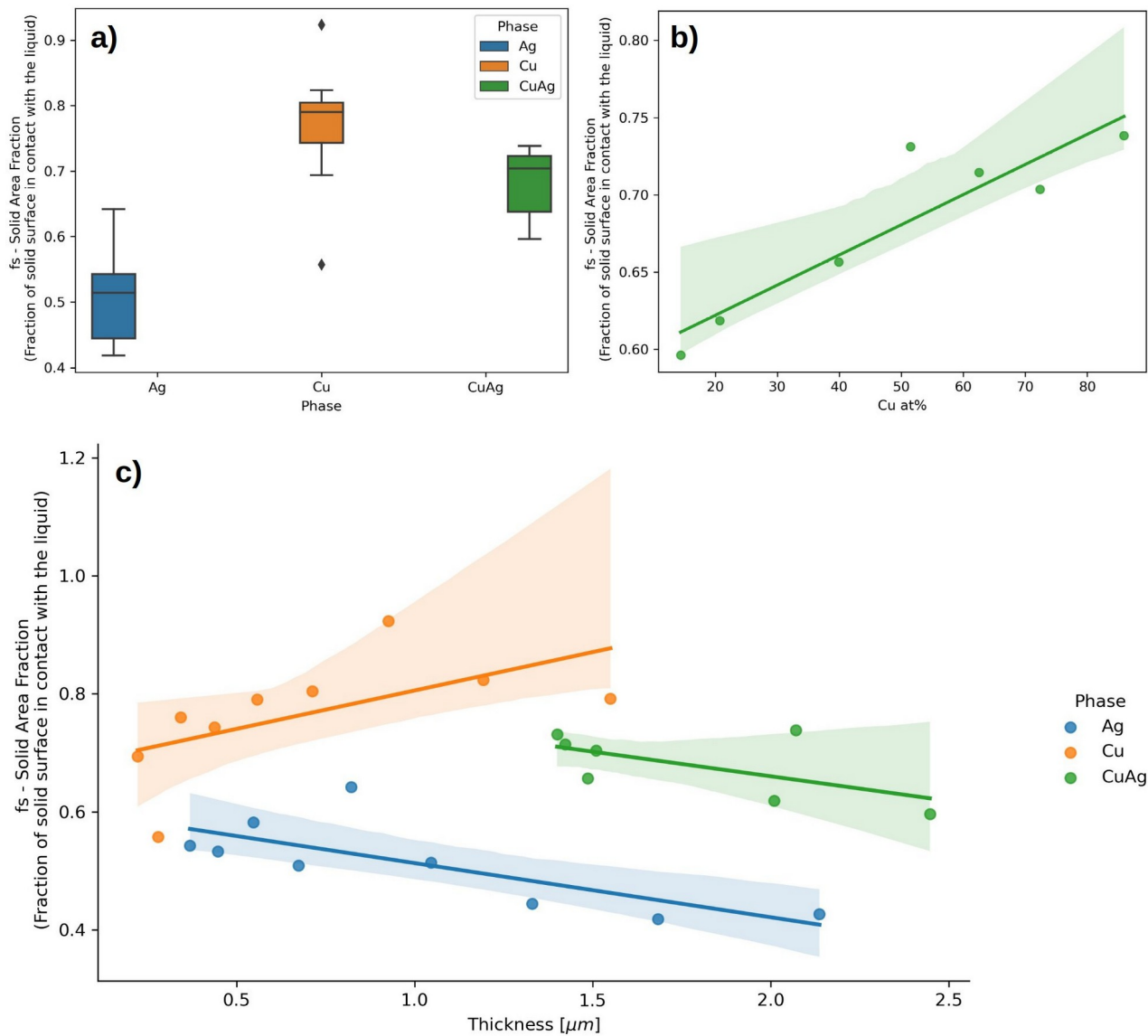

*Figure S11: Derived Solid Area Fraction $f_s$ (Fraction of solid surface in contact with the liquid) assuming Cassie-Baxter wetting state from water contact angle measurements on textured surface for the Ag, Cu and CuAg layers. a) Solid Area Fraction $f_s$ Box Plot for all the layer phases. b) Solid Area Fraction $f_s$ as a function of copper content for CuAg. c) Solid Area Fraction $f_s$ as a function layer thickness.*

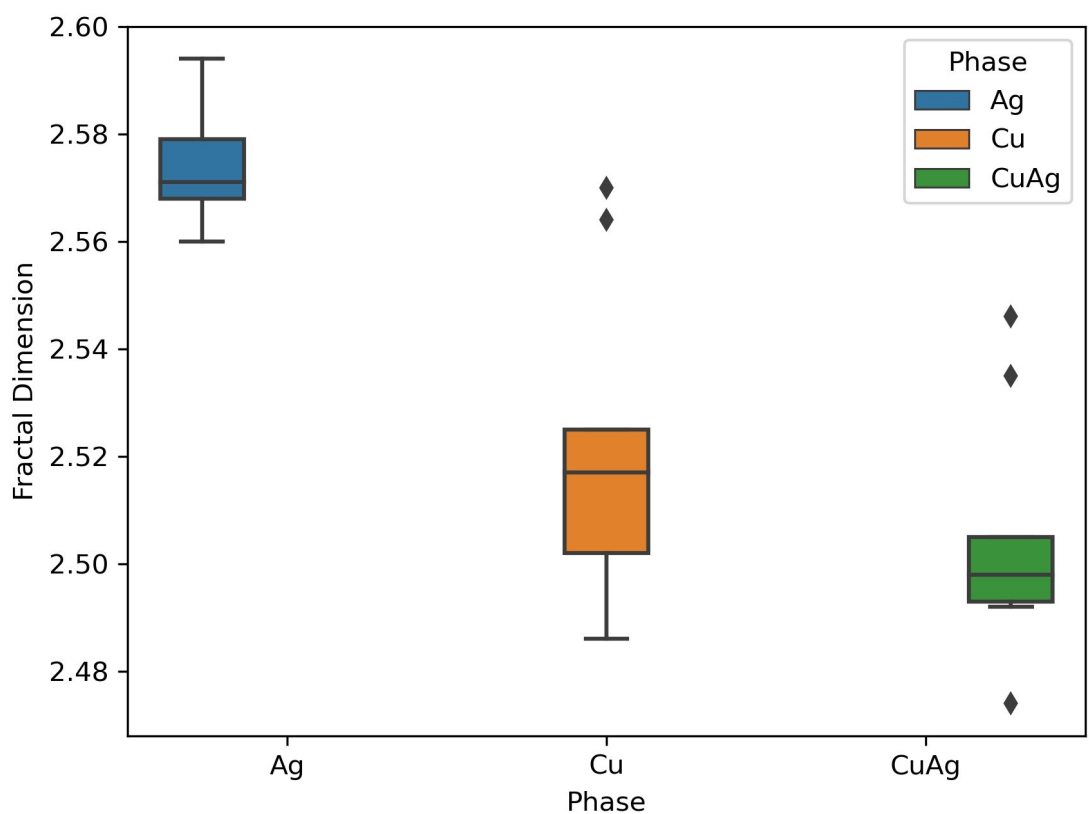

*Figure S12: Surface fractal dimension box plot for all layer phases (Ag, Cu, CuAg). The Ag phase exhibits the highest fractal dimension, indicating a more complex, irregular, and self-similar surface morphology with greater surface asperities and hierarchical micro/nanostructures. In contrast, Cu and CuAg show similar and lower fractal dimensions, reflecting a relatively smoother and less complex surface topography. This increased geometric complexity in Ag correlates with enhanced hydrophobicity, as the multiscale roughness promotes air trapping and stabilizes the Cassie–Baxter wetting regime. The higher fractal dimension of Ag thus contributes to improved water repellency through greater surface heterogeneity and multiscale roughness.*